# Dynamic origin of azimuthal modes splitting in vortex-state magnetic dots

Konstantin Y. Guslienko,[1] Andrei N. Slavin,[2] Vasyl Tiberkevich[2] and Sang-Koog Kim[1*]

[1] *Research Center for Spin Dynamics & Spin-Wave Devices and Nanospinics Laboratory, Department of Materials Science & Engineering, Seoul National University, Seoul 151-744, South Korea*

[2] *Physics Department, Oakland University, Rochester, MI 48309, U.S.A.*

A spin wave theory explaining experimentally observed frequency splitting of dynamical excitations with azimuthal symmetry of a magnetic dot in a vortex ground state is developed. It is shown that this splitting is a result of the dipolar hybridization of three spin wave modes of a dot having azimuthal indices $|m|=1$: two high-frequency azimuthal excitation modes of the in-plane part of the vortex with indices $m = \pm 1$ and a low-frequency $m = +1$ gyrotropic mode describing the translational motion of the vortex core. The analytically calculated magnitude of the frequency splitting is proportional to the ratio of the dot thickness to its radius and quantitatively agrees with the results of time resolved Kerr experiments.

Fundamental understanding of magnetization (**M**) dynamics in geometrically confined magnetic elements is of crucial importance for the future advancements in nanomagnetism and spintronics [1-2]. This understanding can be achieved by combining analytical theory and numerical simulations with advanced time/space-resolved experimental techniques. The **M** dynamics becomes especially interesting and non-trivial in mesoscopic patterned magnetic elements that admit spatially non-uniform ground states, and, in particular, a highly symmetric "vortex" ground state **M(r)** [3-7]. The dynamic spin wave (SW) excitations [8-12] on the background of a vortex ground state govern field- or current- induced reversals of the dot **M** and generation of propagating SWs in the vortex-state magnetic dots [13-14].

The mesoscopic dot ground state in the form of a magnetic vortex, which consists of the in-plane curling **M** and the out-of-plane **M** at the core, is a simplest topologically non-trivial ground state in a ferromagnetic dot [see the inset of Fig. 1(a)]. It has been established both theoretically [7,12,15] and experimentally [9-11, 16-18], that the SW excitation spectrum in the vortex-state dots consists of a low-frequency gyrotropic mode describing the translation (orbital) motion of the vortex core [8] and high-frequency modes with radial and azimuthal symmetry, describing dipolar excitations of the peripheral planar part of the vortex. It has been, also, established both experimentally and numerically [10, 11] that the SW azimuthal modes with indices $m = \pm 1$ (i.e., the modes that propagate in opposite azimuthal directions around the core) demonstrate a substantial frequency splitting of the order of 1 GHz for typical dot parameters. Note that a similar splitting was found in simulations done on asteroid-shaped dots with the anti-vortex ground state [19]. It is clear that this splitting is somehow caused by the interaction of the azimuthal SW modes with the vortex (anti-vortex) core, since it completely vanishes if the small central region of the dot containing the "core" is physically removed [11]. An attempt [12] to explain this splitting as a result of azimuthal SW mode interaction with the static dipolar and exchange fields of the vortex core produced the splitting that is several times smaller than the splitting observed experimentally [10,11]. Thus, at the present time the origin of significant frequency splitting of $|m|=1$ SW modes remains elusive. It is, also, unclear, why the other azimuthal SW modes of the dot with azimuthal indices $|m| \neq 1$ have a much smaller splitting.

In this Letter we develop a simple explanation of the splitting of $|m|=1$ modes that quantitatively agrees with experimental data. We demonstrate that the origin of the large SW azimuthal modes splitting is an interaction of the high-frequency azimuthal SW modes with the low-frequency gyrotropic mode. In other words, the splitting is a result of the *dynamic hybridization* of the azimuthal $|m|=1$ SW modes of the in-plane part of the vortex with the gyrotropic mode of the vortex core, which can be considered as a SW mode having the azimuthal indeces $m$=+1 or $m$= -1. Due to the same azimuthal behavior, the gyrotropic mode can hybridize with the high-frequency $|m|=1$ azimuthal modes and can shift their frequencies. This frequency shift depends on the propagation direction of SW mode, and, since two high-frequency modes propagate in the opposite directions, hybridization leads to the splitting between the frequencies of the modes with indices $m$=+1 and -1. The modes with other azimuthal dependences ($|m| \neq 1$) cannot effectively interact with the gyrotropic mode, and, therefore, their splitting is caused by the interaction with a static dipolar field of the vortex core, which is rather weak at a reasonable distance from the dot center (Fig. 1a).

To put the above-described ideas in an analytic form, we derive equations for the linear **M** excitations of a magnetic dot in the vortex ground state. The **M** dynamics is described in the framework of the Landau-Lifshitz equation of motion $d\mathbf{M}/dt = -\gamma \mathbf{M} \times \mathbf{H}_{eff}$, where the effective field $\mathbf{H}_{eff}[\mathbf{M}] = \mathbf{H}_d[\mathbf{M}] + L_e^2 \nabla^2 \mathbf{M}$ includes contributions from the magnetostatic ($\mathbf{H}_d$) and exchange fields ($L_e = (2A/M_s^2)^{1/2}$ is the exchange length, $M_s$ is the saturation magnetization). The magnetostatic field is non-local $\mathbf{H}_d(\mathfrak{R}) = -4\pi \int d^3\mathfrak{R}' \hat{G}(\mathfrak{R},\mathfrak{R}')\mathbf{M}(\mathfrak{R}')$, $\mathfrak{R} = (\mathbf{r},z)$ is the coordinate within magnetic dot, $\hat{G}$ is the tensor kernel singular at $\mathfrak{R} = \mathfrak{R}'$ [15]. The equation of motion is non-linear and integro-differential. We assume that the dot is thin (**M** has no dependence on the coordinate $z$) and linearize this equation by the substitution $\mathbf{M}(\mathbf{r},t)/M_s = \boldsymbol{\mu}(\mathbf{r}) + \mathbf{m}(\mathbf{r})e^{-i\omega t}$, where $\boldsymbol{\mu}(\mathbf{r})$ describes a non-uniform ground state ($\mathbf{M}_0 = \boldsymbol{\mu} M_s$), $\mathbf{m}(\mathbf{r})$ is the small dynamic excitation ($|\mathbf{m}| \ll |\boldsymbol{\mu}|$, $\boldsymbol{\mu} \cdot \mathbf{m} = 0$), and $\mathbf{r} = (r,\varphi)$. Keeping only linear in $\mathbf{m}(\mathbf{r})$ terms we get the linear integro-differential equation to determine the excitation spectrum:

$$-i\omega \mathbf{m} = \boldsymbol{\mu} \times \hat{L} * \mathbf{m}, \qquad (1)$$

where the linear self-adjoint operator $\hat{L}$ is defined as $\hat{L} * \mathbf{m} = \Omega \mathbf{m} - \gamma \mathbf{H}_{eff}[\mathbf{m}]$, $\Omega = \gamma \mathbf{H}_{eff}[\boldsymbol{\mu}] \cdot \boldsymbol{\mu}$ and has contributions from the dynamic dipolar and exchange fields, and static dipolar field of the ground state. The different spin eigenmodes $\mathbf{m}_\nu$ are orthogonal to each other in the sense

$$\int_{r<R} d^2\mathbf{r}\, \boldsymbol{\mu} \cdot [\mathbf{m}_\nu^* \times \mathbf{m}_{\nu'}] = i N_\nu \delta_{\nu\nu'}, \qquad N_\nu = \frac{1}{\omega_\nu} \int_{r<R} d^2\mathbf{r}\, (\mathbf{m}_\nu^* \cdot \hat{L} * \mathbf{m}_\nu), \qquad (2)$$

where $\omega_\nu$ is eigenfrequency of the $\nu$-th eigenmode, $\mathbf{m}_\nu(\mathbf{r}) = \mathbf{m}_\nu(r)\exp(im\varphi)$, integration is conducted over a cylindrical dot with the radius $R$ and thickness $L$. I.e., the spin eigenmodes $\mathbf{m}_\nu$ satisfy to $-i\omega_\nu \mathbf{m}_\nu = \boldsymbol{\mu} \times \hat{L} * \mathbf{m}_\nu$ and the eigenfrequencies $\omega_\nu$ are calculated as $\omega_\nu = N_\nu^{-1} \langle \mathbf{m}_\nu^* \cdot \hat{L}[\mathbf{m}_\nu] \rangle$, where $N_\nu$ are real normalization constants and $\langle f \rangle = \int f(\mathbf{r}) d^2\mathbf{r}$. The properties (2) are common for linear excitations in any magnetic system. This calculation of the eigenfrequency $\omega_\nu$ is variationally stable with respect to small perturbations of the mode profile $\mathbf{m}_\nu$ allowing simple estimation of the frequencies of the azimuthal modes. Namely, let $\mathbf{m}_R$ be the azimuthal mode of the magnetic *ring* (circular dot without the vortex core) ground state $\boldsymbol{\mu}_R = \hat{\varphi}$ ($\hat{\boldsymbol{\rho}}, \hat{\varphi}, \hat{\mathbf{z}}$) are the unit vectors along the coordinate axes). $\mathbf{m}_R$ as an eigenmode of the *ring* is not orthogonal (in the sense of Eq. (2)) to the gyrotropic mode $\mathbf{m}_G$ of the vortex dot. Thus, to calculate frequency of the azimuthal modes in the *vortex* ground state, we perform standard orthogonalization procedure of the modes $\mathbf{m}_R$ and $\mathbf{m}_G$. The profile of the azimuthal SW modes in the *vortex* ground state is given by $\mathbf{m}_V = \mathbf{m}_R + i N_G^{-1} (\boldsymbol{\mu} \cdot (\mathbf{m}_G^* \times \mathbf{m}_R)) \mathbf{m}_G$, where $N_G$ is the norm (2) of the mode $\mathbf{m}_G$. Using properties of the modes $\mathbf{m}_R$ and $\mathbf{m}_G$ and substituting them to Eq. (1) yield the approximate expression for the frequency of the azimuthal modes in the presence of the vortex core $\omega_V = (\omega_R - \xi \omega_G)/(1 - \xi)$, $\xi = |\boldsymbol{\mu} \cdot (\mathbf{m}_G^* \times \mathbf{m}_R)|^2 / N_G N_R$, where $\omega_R$ is the frequency of degenerated SW modes ($m$=+1/-1) of the ring, $N_R$ is the norm (2) of the SW modes $\mathbf{m}_R$. The coupling constant $\xi \ll 1$ depends on the propagation direction of the azimuthal SW and leads to the approximate frequency splitting of the $m$=+1/-1 azimuthal SW eigenmodes $\Delta\omega = 2\xi\omega_R$. Here we neglected influence of the vortex core in evaluation of the quantities that involve only the mode $\mathbf{m}_R$, i.e. we have neglected interaction of the azimuthal mode with the *static* field of the core. This estimation also neglects dipolar inter-mode interaction, which may be not small.

In general, the vortex excitation spectra consist of the core gyrotropic mode $\mathbf{m}_G$ and high-frequency SW, $\mathbf{m}_j$, $j>0$. To find the spectrum of the excitations within the general linear approach we project Eq. (1) into some system of the vector functions $\mathbf{m}_j(\mathbf{r})$, all of each are orthogonal to the ground state $\mathbf{\mu}(\mathbf{r})$ and satisfy boundary conditions on the dot circumference $\mathbf{m}(\mathbf{r}) = \sum_j c_j \mathbf{m}_j(\mathbf{r})$. Substituting this decomposition to Eq. (1) we get the system of linear equations for the excitation amplitudes $c_j$:

$$\sum_{j'} \left( L_{jj'} - \omega N_{jj'} \right) c_{j'} = 0, \tag{3}$$

where the Hermitian matrix elements are

$$L_{jj'} = \int_{r<R} d^2\mathbf{r} \left( \mathbf{m}_j^* \cdot \hat{L} * \mathbf{m}_{j'} \right), \quad N_{jj'} = -i \int_{r<R} d^2\mathbf{r} \mathbf{\mu} \cdot \left[ \mathbf{m}_j^* \times \mathbf{m}_{j'} \right]. \tag{4}$$

The vortex ground state is known $\mathbf{\mu} = (0, \mu_\varphi, \mu_z)$, and $\mu_z(0) = p = \pm 1$ is the core polarization [8]. I.e., the problem is reduced to choice of the approximate functions $\mathbf{m}_j(\mathbf{r})$. To find the magnitude of the frequency splitting it is sufficient to take into account only three modes: "unperturbed" gyrotropic mode of the vortex core and ring azimuthal modes with $m=+1/-1$. Thus, the correct eigenfrequencies of the dot can be found as a solution of the cubic secular equation that describes hybridization and dynamic dipolar interaction between the "unperturbed" diagonal modes. Bellow we describe details of the calculations of the vortex gyrotropic mode and the unperturbed SW for a ring.

The gyrotropic mode has the azimuthal index $m=p$. The most convenient language to describe this mode is using the core coordinate $\mathbf{X}(t)$. To include this mode to the standard linear scheme we decompose the shifted vortex $\mathbf{M}$ (the reduced shift $\mathbf{s} = \mathbf{X}/R$ is $\ll 1$) as $\boldsymbol{\mu}(\mathbf{r},\mathbf{s}) = \boldsymbol{\mu}(\mathbf{r},0) + (\mathbf{s}\cdot\partial/\partial\mathbf{s})\boldsymbol{\mu} + O(s^2) = \boldsymbol{\mu}(\mathbf{r},0) + \text{Re}[\bar{s}(\partial\boldsymbol{\mu}/\partial s_x + i\partial\boldsymbol{\mu}/\partial s_y)]$. Thus, we can define the profile of the gyrotropic mode ($j=0$) as

$$\mathbf{m}_G(\mathbf{r}) = \frac{\partial \boldsymbol{\mu}(\mathbf{r},\mathbf{s})}{\partial s_x} + i \frac{\partial \boldsymbol{\mu}(\mathbf{r},\mathbf{s})}{\partial s_y}, \qquad (5)$$

where derivatives are calculated at $s = s_x + i s_y \to 0$. The circular gyrotropic motion $s(t) = c_0 \exp(i\omega_R t)$ is in the counter-clockwise/clockwise direction for $p=+1/-1$. The mode eigenfrequency is $\omega_G = L_{00}/|N_{00}|$. The moving vortex generates some perturbed magnetization outside the core and related dynamical dipolar field. To find $\mathbf{m}_G$ explicitly we use decomposition $\boldsymbol{\mu}(\mathbf{r},\mathbf{s}) = \boldsymbol{\mu}(\mathbf{r},0) + C\boldsymbol{\rho}(1/\rho^2 - 1)(xs_y - ys_x)/\rho$ corresponding to the strong pinning conditions at the dot border [7, 20]. The definition (5) allows writing outside the core $m_G^\rho = iC\rho(1/\rho^2 - 1)\exp(i\varphi)$, $\rho = r/R$ within the pole-free model of the shifted vortex (Fig. 1a). $m_G^\rho \sim 1/\rho$ decreases slowly, whereas $H_d^z \approx 0$ at $r > R_c$. The component $m_G^z$ depends on detailed structure of the moving core and cannot be so simply calculated. But in the main approximation it can be neglected because it has significant values only within the core having very small radius $R_c \sim 10$ nm [4] in comparison to $R \sim 1$ μm.

The spin waves ($j>0$) can be described by eigenfunctions, which depend on two indices, $n$– radial number and $m$-azimuthal number: $\mathbf{m}_\nu(\mathbf{r}) = \mathbf{m}_\nu(r)\exp(im\varphi)$, where $\nu = (n,m)$. Neglecting the exchange interaction the SW eigenfrequencies/eigenfunctions can be found from solution of the linear integral equation (using only $\rho\rho$-component $G(\rho,\rho') = \partial^2/\partial\rho\partial\rho' \int_0^\infty dk k^{-1} f(\beta k) J_1(k\rho) J_1(k\rho')$ of the tensor $\hat{G}$ in the main approximation, $m=+1/-1$, and presenting frequency in the units of $\omega_M = \gamma 4\pi M_s$):

$$\omega^2 m(\rho) = \int_0^1 d\rho' \rho' G(\rho,\rho') m(\rho'). \tag{6}$$

Eq. (6) can be solved decomposing the eigenfunctions in series of the Bessel functions $m(\rho) = \sum_l b_l J_1(\alpha_{1l} \rho)$, where $\alpha_{1l}$ are the roots of the equation $J_1(x)=0$ (see Fig. 1b). The first order Bessel functions $J_1(x)$ were chosen here because they represent solution of the problem for infinitely thin disk (accounting the exchange interaction and simplified magnetostatics) and satisfy the strong pinning boundary conditions at the dot circumference $r=R$.

We apply the above described procedure to calculate the frequency splitting of the SW main modes with $m=\pm 1$ and $n=0$ (no radial zeros) over the vortex ground state. But the scheme can be applied for arbitrary $n$. We assume that the splitting is caused by interaction of the SW $m=\pm 1$ and gyrotropic modes and account only three excitation modes ($j=0,1,2$) with $|m|=1$: the unperturbed gyrotropic mode and two azimuthal SW modes with right/left polarizations $\mathbf{m}_{1,2}(\mathbf{r})$. We use the ansatz $\mathbf{m}_{1,2}(\mathbf{r}) = (\hat{\boldsymbol{\rho}} \pm i\hat{\mathbf{z}}) m(\rho) \exp(i\varphi)$, where $m(\rho)$ is some radial mode profile calculated from Eq. (6). These assumptions allow us to simplify Eq. (3) using the symmetry relations: $L_{11} = L_{22}$, $L_{21} = L_{12}$, $L_{01} = L_{02}$, $N_{22} = -N_{11}$, $N_{12} = N_{21} = 0$, and $N_{02} = -N_{01}$. The expression for the diagonal SW frequencies (no interaction with the gyrotropic mode) $\omega_R^2 = (L_{11}^2 - L_{12}^2)/N_{11}^2$ is similar to one for the radial vortex modes

[15]. We get from Eq. (3) the secular equation:

$$\det(\hat{L} - \omega\hat{N}) = (\omega_G - \omega)(\omega_R^2 - \omega^2) + \frac{N_v^2}{2N_s}\left[\omega^2(\omega_R^2 - 2\omega_D^2) + \omega_D^4\right] = 0, \quad (7)$$

where $N_v = \int d\rho(1-\rho^2)m(\rho)$ is overlapping of the gyrotropic and azimuthal $m(\rho)$ SW modes, $N_s = \int d\rho \rho m^2(\rho)$, $\omega_G = (1/2)\int_0^\infty dk\, k^{-1} f(\beta k) I^2(k)$, $\omega_D^2 = N_v^{-1}\int_0^\infty dk\, k^{-1} f(\beta k) I(k) I_s(k)$ describes the inter-mode dipolar coupling, $I(k) = 2\int_0^1 d\rho \rho J_1(k\rho)$, $I_s(k) = \int_0^1 d\rho \rho m(\rho) \partial J_1(k\rho)/\partial \rho$, and $f(x) = 1 - (1-\exp(-x))/x$.

The azimuthal SW frequencies increase approximately as $\sim \sqrt{L/R}$ [15]. In the main approximation the azimuthal SW frequencies (Fig. 2a) and their splitting (Fig. 2b) can be calculated from Eq. (7) to be

$$\Delta\omega = \frac{N_v^2}{2N_s}(\omega_R^2 - 2\omega_D^2). \quad (8)$$

The frequency splitting (8) is $\Delta\omega \approx 3.5\omega_G$. The approach allows to calculate $\xi = \omega_R N_v^2/4N_s$ yielding the approximate splitting $\Delta\omega = (N_v^2/2N_s)\omega_R^2$. This expression neglects the inter-mode dipolar coupling and leads to overestimation of the frequency splitting in ~ 15-20 % in comparison to Eq. (8).

Summarizing, in our model all the dynamic eigenmodes (radial, azimuthal, and gyrotropic) of the vortex state magnetic dot are described within an unified perturbative approach, and the frequency splitting of the azimuthal SW modes with the azimuthal indices $m=\pm 1$ is interpreted as a hybridization of these modes with the gyrotropic mode. The SW mode rotating in the direction of the gyrotropic mode decreases its frequency, whereas one rotating in the opposite direction increases it. The azimuthal modes splitting caused by their interaction with the static dipolar field of the vortex core [12] also exists, but it is one order of magnitude smaller than the dynamical splitting for $|m|=1$ described by our theory. The hybridization of the gyrotropic and azimuthal SW modes leads, in particular, to distortion of the vortex core profile and can be considered as first stage of the core *p*-reversal [13, 14]. The maximum amplitude of SW rotating in the same direction as the gyrotropic mode approaches the core trajectory and the maximum of oppositely rotating SW shifts apart from the dot center. The splitting reverses its sign if the vortex polarization reverses because the core velocity changes its sign. The frequency splitting calculated by Eq. (8) accounting the SW profiles determined by Eq. (6) can be compared to one measured in Refs. 10, 11. The calculated frequency splitting increases proportionally to the dot aspect ratio (thickness/radius), and its magnitude is of ~ 1-2 GHz in good agreement with the results of the recent time-resolved Kerr measurements [10, 11] of permalloy dots.

We thank to P.A. Crowell for supplying us the original experimental data. This work was supported by the Creative Research Initiatives (ReC-SDSW) of MEST/KOSEF, as well as by the MURI grant W911NF-04-1-0247 from the Department of Defense of the USA, by the grant W911NF-04-1-0299 from the U.S. Army Research Office, by the contract No.W56HZV-07-P-L612 from the U.S. Army TARDEC, RDECOM, by the grant ECCS-0653901 from the National Science Foundation of the USA, and by the Oakland University Foundation.

# References

*Corresponding author, electronic address: sangkoog@snu.ac.kr

**Figure captions**

**FIG. 1** (color online). The unperturbed eigenfunctions of the vortex gyrotropic motion, $z$-component of the dipolar field from the static core (the dashed line) in (a) and the azimuthal spin wave eigenmode profiles with indices $m=+1/-1$ and $n=0$ (solid), 1 (dashed), 2 (dot-dashed lines) in (b). Inset of (a) shows cylindrical magnetic dot in the vortex state. The spike in the center is the vortex core with positive polarization and the in-plane curling **M** distribution is marked by different colors.

**FIG. 2** (color online). The splitted frequencies of the vortex azimuthal SW modes with indices $m=\pm1$, $n=0$ (a) and the value of the splitting $\Delta\omega$ (b) according to Eq. (8) plotted along with the vortex gyrotropic eigenfrequency $\omega_G$ (dashed line) vs. the dot aspect ratio with (solid line) and without (dotted line) accounting the inter-mode dipolar coupling. The symbols correspond to the experimental data of Refs. 10 and 11 (the splitting 1.25 GHz). $M_s$= 770 G, $\gamma$= 2.95 MHz/Oe.

**FIG. 1**

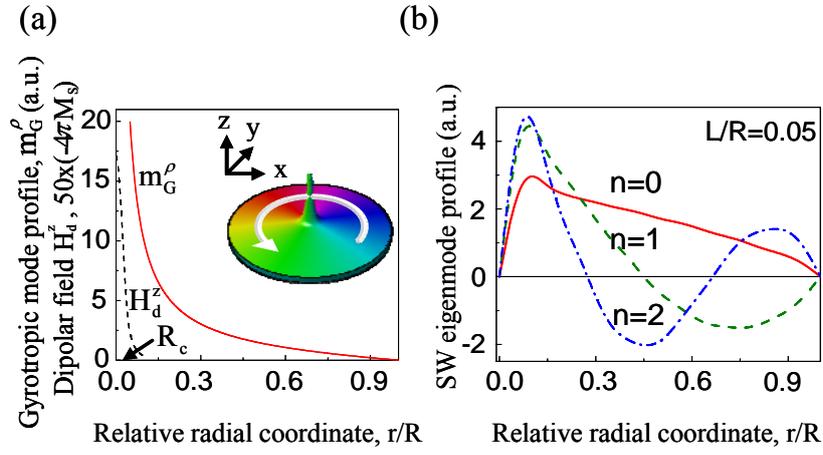

**FIG. 2**

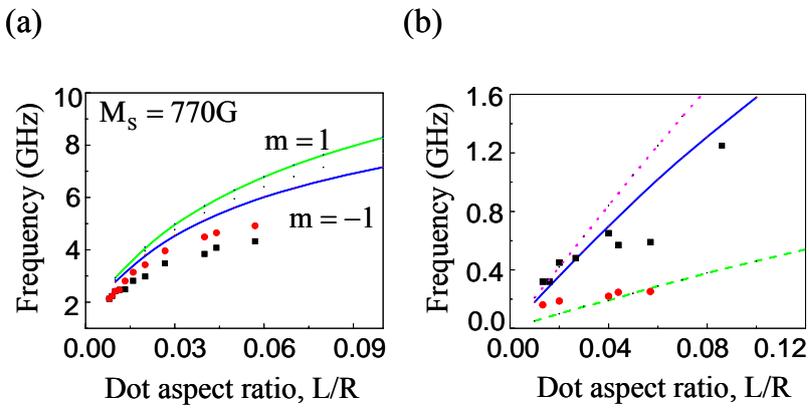